\documentclass[aps,prb,reprint,superscriptaddress]{revtex4-1}
\usepackage{graphicx,amsfonts,amsmath,latexsym,amssymb}

\newcommand{\mags}[0]{\mu_\sigma}
\newcommand{\msd}[0]{\mu_{-\sigma}}
\newcommand{\nums}[0]{n_\sigma}

\newcommand{\dms}[0]{\partial_T \mu_\sigma}
\newcommand{\dmsd}[0]{\partial_T \mu_{-\sigma}}
\newcommand{\dns}[0]{\partial_T n_\sigma}
\newcommand{\dnsd}[0]{\partial_T n_{-\sigma}}

\newcommand{\mua}[0]{\mu_\uparrow}
\newcommand{\nua}[0]{n_\uparrow}

\newcommand{\dmua}[0]{\partial_T \mu_\uparrow}
\newcommand{\dnua}[0]{\partial_T n_\uparrow}

\newcommand{\mda}[0]{\mu_\downarrow}
\newcommand{\nda}[0]{n_\downarrow}

\newcommand{\dmda}[0]{\partial_T \mu_\downarrow}
\newcommand{\dnda}[0]{\partial_T n_\downarrow}

\newcommand{\dos}[0]{\rho\left(\epsilon\right)}

\newcommand{\lns}[0]{\ln \left(1+e^{-\frac{\epsilon-\mu_\sigma}{k_B T}}\right)}

\newcommand{\sr}[0]{Sr$_3$Ru$_2$O$_7$}

\begin{document}

\title{The role of bandstructure in the thermodynamic properties of itinerant metamagnets}
\author{A.M. Berridge}
\affiliation{School of Physics and Astronomy, University of Birmingham, Edgbaston, Birmingham, B15 2TT United Kingdom}
\affiliation{School of Physics and Astronomy, University of St. Andrews, North Haugh, St. Andrews KY16 9SS, United Kingdom}
\date{\today}

\begin{abstract}
It is known that itinerant metamagnetic transitions can be driven by features in the electronic density of states.  We study the signatures of these transitions in the entropy and specific heat for a variety of different cases, identifying the key features which differ from naive expectations, such as enhanced critical fields and non-Fermi liquid temperature dependencies.  We begin with the generic case of a logarithmically divergent density of states, as caused by a two dimensional van Hove singularity.  We then study a specific model for the bandstructure of Sr$_3$Ru$_2$O$_7$, a material with a well-studied metamagnetic transition and quantum critical endpoint.  We consider how far the behaviour of the system can be explained by the density of states rather than quantum fluctuations, and the distinctive features of this mechanism.  One of the characteristic features of Sr$_3$Ru$_2$O$_7$ is an unusual phase with a higher entropy than its surroundings, we consider how this may arise in the context of a density of states picture and find that we can reproduce the thermodynamic behaviour and first-order phase transitions.
\end{abstract}

\maketitle

There has been much interest in the magnetic phase diagrams of itinerant electron systems.  These show both thermal and quantum phase transitions and often non-Fermi liquid behaviour associated with quantum critical (end)points.  A particularly well studied example is \sr.  This displays a metamagnetic transition which bifurcates as a function of field angle to enclose an anomalous phase where the transport properties break the symmetry of the crystal lattice~\cite{327_Science2004,327_Science2007}.  Intriguingly this phase has been shown to have a higher entropy than the surrounding `normal' phases~\cite{327_Science2009}, contrary to naive expectations.  The metamagnetism and anomalous phase are generally thought to be caused by the presence of a van Hove singularity just below the Fermi surface in one of the electronic bands of the material~\cite{327_PRL2008}.  Such peaks in the density of states have been shown to reproduce the metamagnetic transition \cite{Binz04,Theory_magFFLO,Theory_magFFLO_2}.  However, there is known to be a quantum critical endpoint in the region of the anomalous phase~\cite{327_Science2001}, and the possible role of this critical point in the formation of the phase remains largely unexplored.  In addition the region around the phase shows signatures which may be attributed to the quantum critical point, such as a diverging entropy and specific heat.  Also the dependence of the metamagnetic transition on field angle, as well as doping~\cite{doping} and STM~\cite{STM} studies, have cast doubt on the simple picture of a fixed bandstructure with the field acting to Zeeman split the spin-species through a peak in the density of states.

These issues raise the question of how far the properties of the material may be explained by the density of states, without involving quantum fluctuations or more exotic physics.  We address this question by studying the evolution of an itinerant magnetic system as a function of applied magnetic field and temperature where the density of states has a peak or other sharp feature near to the Fermi surface.  In particular we will focus on the entropy and specific heat of the system near to a metamagnetic transition.  We will study both the generic logarithmic divergence of the density of states for a van Hove singularity in a two dimensional system and a simple model for the electronic band in \sr\ which contains the singularity.  We find that as well as the main metamagnetic transition such a model provides a natural explanation for the magnetic crossover observed at slightly lower fields and connects this with features observed in the entropy and specific heat~\cite{327_Science2009}.  The peak produces a logarithmic divergence of $C/T$ with temperature and unusual behaviour as a function of field even in the absence of a quantum critical point.  We will consider how these results may help to identify the cause of the metamagnetic transition in \sr\ and may more generally help to distinguish between density of states features and quantum critical effects.  

We will go on to consider a toy model for the entropic properties of the anomalous phase of \sr.  One of the striking features of this phase is that it has a higher entropy than the surrounding regions, giving rise to phase transitions which `fan out' as temperature is increased.  We will show that a density of states with a closely spaced double-peak will reproduce this phase diagram and thermodynamic behaviour and speculate on the origin of such a feature in the density of states. 

We perform calculations of the magnetisation, specific heat and entropy from the mean-field Stoner model for a general density of states.  We take care to include the condition of number conservation.  These results must be evaluated numerically for any given bandstructure.  Having identified the mechanisms which contribute to these results we present the numerical evaluation of the calculations for a cut through the metamagnetic wing of the phase diagram.   We will begin with the generic case of a logarithmic singularity in the density of states, caused by a saddle point in the electronic dispersion in two dimensions~\cite{vanHove}, which has been shown to induce a metamagnetic transition~\cite{Binz04}.  We will then go on to study the case of a model bandstructure for the $\gamma_2$ band of \sr ~\cite{327_PRL2008} which has been shown to contain the van Hove singularity.  This shows many of the same features as the logarithmic case with the addition of a magnetic crossover.  Finally we consider a double-peak density of states which gives an interesting entropic behaviour, with two metamagnetic transitions and a high entropy plateau region between them.

\section{Heuristic discussion}

The effects contributing to the form of the entropy and specific heat were considered in detail in Ref.\onlinecite{CS1}, we will briefly summarize these results here.

The shape of the entropy and specific heat curves for an itinerant magnet may be deduced from some basic principles.  Entropy as a function of field should follow the density of states as a function of energy.  There are several effects which alter this dependence in the presence of a varying density of states.  These include spin-splitting due to the external magnetic field, interaction induced magnetism, number conservation~\cite{confproc,327_JF_PRL} and temperature.

\begin{figure*}
\centering
\includegraphics[width=6.5in]{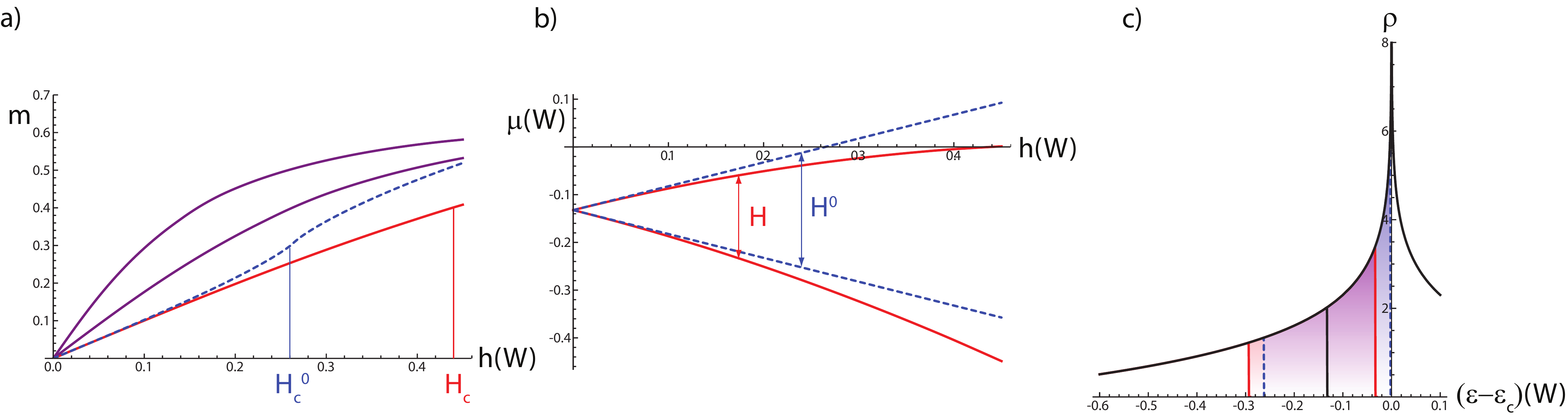}
\caption{\label{fig:numberconservation}(Color online) a) Magnetisation for the number conserving case (red solid line) and non-number conserving case (blue dashed line, determined by the equation $2m=n(\epsilon-\mu-h/2)-n(\epsilon-\mu+h/2)$, valid in the case of a non-interacting system).  The critical field $H_c$ is higher in the number conserving case than the non-number conserving case $H_c^0$.  Introducing interactions (purple solid lines, $g=0.22, 0.35$) reduces the critical field.  $W$ is the bandwidth.  b) Effective chemical potentials relative to the singularity for the two spin species, with number conservation (solid red) and non number conserving (dashed blue).  The number conserving case slows its approach to the van Hove singularity as field is increased.  c)  The density of states with the zero-field chemical potential (solid black line) the non number conserving chemical potentials at the critical field (blue dashed line) and the number conserving case at the same field (solid red lines).}
\end{figure*}

Upon application of a magnetic field the spin-species' Fermi surfaces become split.  The entropy is then given by the sum of the density of states at two different energies.  Since one Fermi surface is moved to a lower density of states this has the effect of compressing the peak in entropy around the van Hove singularity.  This splitting is proportional to the magnetic field.  However, by moving a fixed energy interval the Fermi surface closest to the peak expands to include more electrons than the Fermi surface further from the peak loses by contracting (assuming the Fermi surfaces lie below the peak).  This means that the overall number of electrons has increased.  In order to conserve number the majority Fermi surface must move towards the peak more slowly than the minority Fermi surface recedes.  This results in a slower approach to the peak than would naively be expected and therefore a higher critical field for the metamagnetic transition.  This is illustrated in detail in figure \ref{fig:numberconservation}.  We note that despite this effect the energy gap between the Fermi surfaces remains proportional to the field.
 
When interactions are included as well as the external field there is an additional splitting of the Fermi surfaces due to metamagnetism.  When the transition is continuous the effect is one of compressing the field scale around the transition so that field tunes through the putative entropy curve more rapidly than expected (see Fig.\ref{fig:numberconservation}).  When the transition is discontinuous a range of magnetisation values are `jumped over' by the transition.  This removes a slice of the putative entropy curve beginning at the critical endpoint and getting wider as temperature is decreased and the transition gets stronger.  The entropy therefore becomes discontinuous at the metamagnetic transition.  Since the region removed is around the peak, it is possible for the highest value of $S/T$ to occur at non-zero temperature, where the jump is smaller and the Fermi surface samples a region closer to the peak.  This is similar to the susceptibility which is strongly peaked around the critical endpoint of the transition.

Finally we consider the effect of temperature.  This broadens the Fermi-Dirac distribution, allowing the thermal occupation the peak in the density of states across a wider range of chemical potentials as temperature is increased.  The peak in entropy therefore becomes broadened.  The increase in occupied states is most rapid on either side of the peak.  Specific heat is the temperature derivative of the entropy and so will be largest in these regions.  We therefore expect a structure which is logarithmic at zero temperature with a peak which bifurcates and broadens as temperature increases.

The magnetic field and interactions have the same effect for the specific heat as for the entropy - a compression along the field axis and the removal of a `wedge' of field values due to the first order transition.

We will now consider how to calculate these quantities explicitly within the Stoner mean-field theory.

\section{Calculation of entropy and specific heat}
\label{deriv}

We calculate the magnetization, entropy and specific heat from the free energy of the Stoner model where the effective chemical potential for the spin-species is determined by number conservation.  Number conservation is enforced by requiring $n_\sigma=\frac{n}{2}+\sigma m$ where $\sigma=\pm1$ labels the spin-species, the total number of electrons $n=n_\uparrow+n_\downarrow$ is constant, and the magnetisation $2m=n_\uparrow-n_\downarrow$.  The effective chemical potential $\mu_\sigma$ is determined implicitly from the number $n_\sigma$,
\begin{eqnarray}
n_\sigma=\int d\epsilon \ \rho(\epsilon) f(\epsilon-\mu_\sigma),
\end{eqnarray}
where $f(\epsilon-\mu_\sigma)=\left[1+\exp\left\{\frac{1}{k_B T}\left(\epsilon-\mu\right)\right\}\right]^{-1}$ is the Fermi-Dirac distribution and $\rho(\epsilon)$ is the density of states.  Throughout all integrals are performed over the bandwidth of the band in question.

The free energy in the Stoner model is:
\begin{eqnarray}
F
&=&
\sum_\sigma \left[ -k_B T \int d\epsilon \ \dos \lns + \mags \nums \right]
\nonumber\\
&& + g \nua \nda -hm,
\label{fenergy}
\end{eqnarray}
where $g$ is the interaction strength and $h$ is the applied magnetic field multiplied by $\mu_B$.  From the requirement that the free energy is a minimum, $\partial_m F=0$, this gives a self-consistent equation for the magnetization and magnetic susceptibility~\cite{Binz04},
\begin{eqnarray}
h
&=&
\mua(n, m)-\mda(n, m)-2gm,
\label{tm}
\end{eqnarray}
\begin{eqnarray}
\frac{1}{\chi}
&=&
\sum_\sigma \frac{1}{\int d\epsilon \ \rho\left(\epsilon\right) \partial_\epsilon f\left(\epsilon-\mu_\sigma\right)} - 2g.
\label{sus}
\end{eqnarray}
The entropy is defined by $S=-\left.\partial_T F \right|_{n, h}$ and the specific heat as $C=-T\left.\partial_T^2 F\right|_{n, h}$.  These are evaluated with the conditions that total number is conserved.  This condition is encoded in the behaviour of the chemical potentials, giving a non-trivial form for $\partial_T \mu_\sigma$.  The evaluation of these derivatives is straightforward but lengthy and is presented in appendix \ref{thermoderiv}.  The results of these calculations are:
\begin{eqnarray}
S
&=&
\sum_\sigma \left[k_B \int d\epsilon \ \dos \lns \right.
\nonumber\\
&&+ \left. k_B T\int d\epsilon \ \dos \partial_T \lns - \dms\nums \right],
\nonumber\\
C
&=&
\sum_\sigma \left[\int d\epsilon \ \epsilon \dos \partial_T f\left(\epsilon-\mu_\sigma\right)\right]
-
\left(2gm+h\right)\dnua.
\nonumber\\
\label{sac}
\end{eqnarray}
The temperature derivatives of the chemical potential are given by
\begin{eqnarray}
\dms
&=&
\frac{
-k_B T\frac{\int d\epsilon \; \left(\Xi_\downarrow \frac{\epsilon-\mda}{k_B T^2}+\Xi_\uparrow \frac{\epsilon-\mua}{k_B T^2}\right)}{\int d\epsilon \; \Xi_{(-\sigma)}}
+
2g \int d\epsilon \; \Xi_\sigma \frac{\epsilon-\mags}{k_B T^2}}
{1-\frac{2g}{k_B T}\int d\epsilon \; \Xi_{\sigma}+\frac{\int d\epsilon \; \Xi_{\sigma}}{\int d\epsilon \ \Xi_{(-\sigma})}},
\nonumber\\
\label{dmu}
\end{eqnarray}
where
\begin{eqnarray}
\Xi_\sigma
&=&
\rho(\epsilon)
\frac{e^{\frac{(\epsilon-\mu_\sigma)}{k_B T}}}{\left(1+e^{\frac{(\epsilon-\mu_\sigma)}{k_B T}}\right)^2}.
\end{eqnarray}

The specific heat and entropy may be calculated for any $n$, $h$ and $T$ from equations (\ref{tm}),(\ref{sac}) and (\ref{dmu}).  These expressions produce the magnetic transitions of the Stoner model, although the location of the first-order transition must be determined by direct minimisation of the free energy as it is not uniquely determined by Eq.\ref{tm} which becomes multi-valued in the region of the first-order transition.  We will evaluate these expressions for magnetisation, entropy and specific heat numerically for the logarithmic density of states in section \ref{sec:results}, the model $\gamma_2$ band in section \ref{sec:g2} and a toy model double-peaked density of states in section \ref{sec:double}.

\section{Density of states with logarithmic peak}
\label{sec:results}

With the previous results it is possible to evaluate the magnetization, susceptibility, entropy and specific heat as a function of filling, magnetic field and temperature, for any given density of states and interaction strength.  First we will study a logarithmically divergent density of states, as produced generically by saddle points in a two dimensional electronic dispersion.  This model density of states is given by
\begin{eqnarray}
\dos
&=&
\frac{1}{W} \ln\left|\frac{W}{\epsilon-\epsilon_c}\right|
\label{LogDos}
\end{eqnarray}
where the bandwidth is $2W$.  The density of states diverges at $\epsilon=\epsilon_c$. In the following we will take the interaction strength to be $g=0.3W$.  We choose to look at a filling which is below the van Hove point and use field to tune the system through the metamagnetic transition.  The phase diagram for this model was studied in Ref.\onlinecite{Binz04} and the thermodynamic properties were briefly studied in Ref.\onlinecite{CS1}, we now consider them in more detail.

\begin{figure*}
\centering
\includegraphics[width=5.25in]{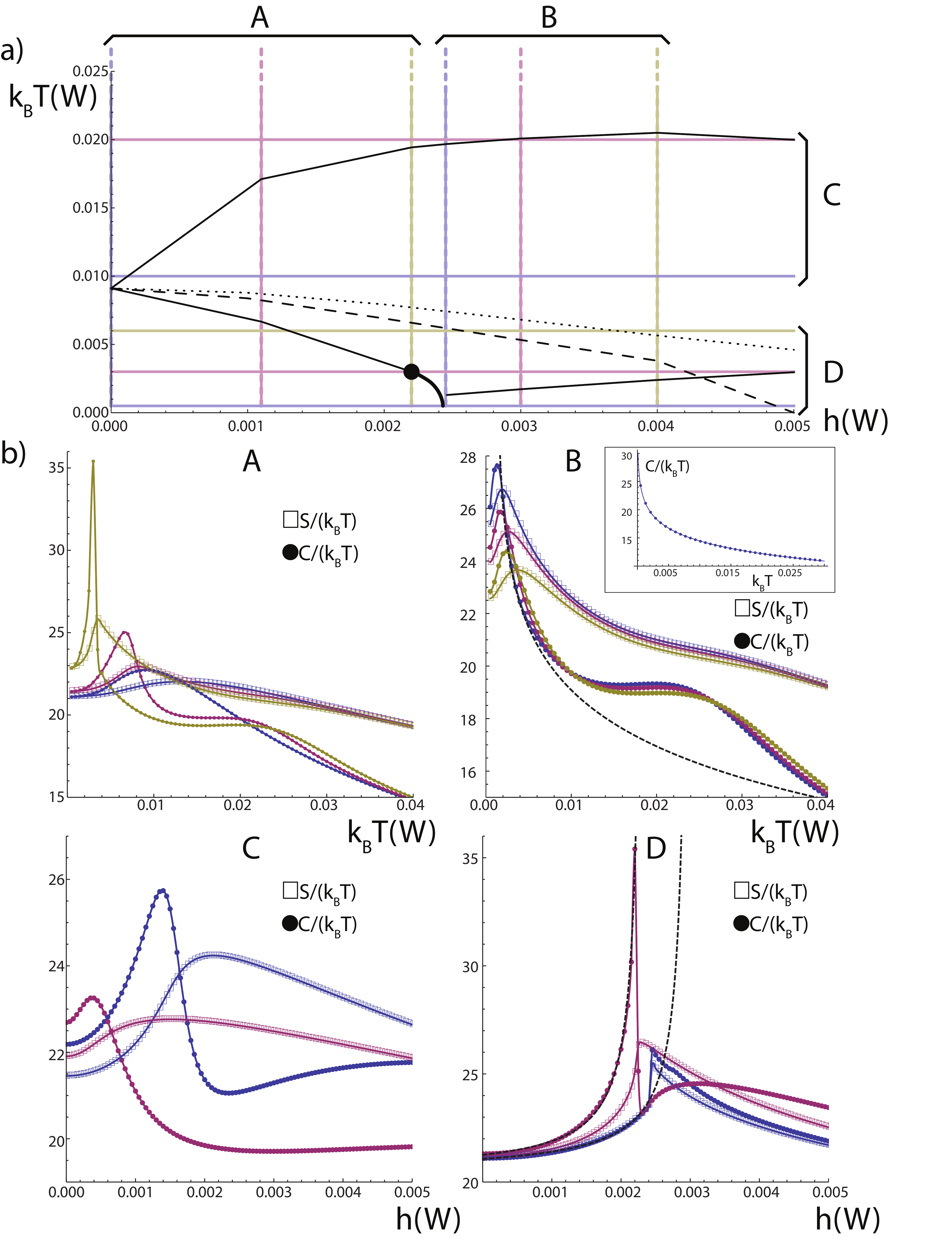}
\caption{\label{fig:specheat}(Color online) a) Phase diagram as a function of magnetic field $h$ and temperature $T$ for $g=0.3W$ and filling fraction $n=0.42$.  The first-order transition is shown by the thick black line, terminating at a critical endpoint.  The thin black lines indicate two peaks in $C/T$ at low temperature while the upper branch follows the shoulder feature.  The dashed line indicates the peak in the specific heat with the interaction strength reduced to $g=0.29$ and the dotted line $g=0.28$.  We see that the critical field of the transition is increased as the interaction strength decreases.  Vertical and horizontal lines labelled A, B, C and D indicate paths through the phase diagram.  b)  Entropy and specific heat plotted along the paths in the phase diagram indicated in a).  A: $C/T$ (circles) and $S/T$ (squares) as a function of temperature for fields below the critical field of the transition.  Note how the single peak at $H=0$ develops a shoulder as field is increased, caused by spin-splitting of the Fermi-surface, while the main peak sharpens and moves to lower temperatures.  B: $C/T$ (circles) and $S/T$ (squares) as a function of temperature for fields above the critical field of the transition.  The dotted line shows a logarithmic fit to the low temperature $C/T$.  The inset shows $C/T$ for a filling far from the van Hove singularity.  The high field necessary to reach the transition means that the contribution from the minority spin is negligible and $C/T$ follows a perfect logarithmic fit of the form $a+b\log{T}$.  C: $C/T$ and $S/T$ as a function of field above the critical temperature of the transition.  Here we see the broadening of the low temperature peaks and the asymmetric double peak in $C/T$.  D: $C/T$ and $S/T$ as a function of field below the critical temperature of the transition.  We see the jump in entropy and specific heat at the first-order transition which is reflected in the curvature of the transition line seen in a).  We also see the appearance of the double peak in $C/T$.  Dotted lines show $h^{-1}$ fits to the low-field $C/T$.}
\end{figure*}

Figure \ref{fig:specheat} gives the results of evaluating (\ref{tm}) for magnetization and (\ref{sac}) for entropy and specific heat, for a cut through the metamagnetic wing in the $h$, $T$ plane with the logarithmic density of states (\ref{LogDos}).  These plots are in good agreement with the anticipated results.  Magnetization has the familiar first-order transition at low temperature which becomes continuous at a critical endpoint.   The entropy has a temperature-broadened peak reflecting the density of states with the position and symmetry of the peak shifting due to the effect of the metamagnetism and number conservation.  Specific heat shows the expected double-peak structure but with very asymmetric peaks due to the magnetic transition.  Other signatures of the metamagnetic transition are observed in the field dependence, such as the discontinuity and maximum of $S/T$ at non-zero temperature.  The slope of the transition to lower fields as the temperature in increased is consistent with the entropy jump at the transition by the Clausius-Clapeyron relation.  The effects discussed modify the field dependence of these quantities from the logarithmic density of states.  In Fig.(\ref{fig:specheat}) we show fits to the specific heat below the transition which show an $h^{-1}$ dependence which is consistent with that observed in the experiments~\cite{327_Science2009}.  

The fact that the Fermi surfaces are spin-split means that as temperature is increased each will intersect with the peak in the density of states at a different temperature.  However the effect of temperature in broadening the transitions means that rather than a double-peak structure the shape of the peak is modified by the presence of a broad background from the Fermi surface furthest from the peak.  This gives a shoulder at high temperature which appears as the field is increased, this is clearly visible in figure \ref{fig:specheat}.  This shoulder will become less pronounced for larger critical fields as a greater temperature broadening will be required for the minority species to see the peak.  This is a distinctive signal of Zeeman splitting near to a feature in the density of states~\cite{*[{}][{ private communication.}] pc_andreas}.

As shown in figure \ref{fig:specheat} the specific heat has a logarithmic dependence on temperature above the metamagnetic transition, reflecting the density of states.  This may be seen from a simple calculation.  Examining the integrals involved in the entropy and specific heat we see that they are based on the Fermi-Dirac distribution and its temperature derivatives.  Approximating the Fermi-Dirac distribution as $f(\epsilon-\mu)=\frac{1}{2}-\frac{\epsilon-\mu}{4T}$ and using the logarithmic density of states $\rho(\epsilon)=\ln \frac{1}{|\epsilon|}$ we see that $C/T \approx \left. \int_{-T}^{T} d\epsilon \ \epsilon \rho(\epsilon)  \partial_T f(\epsilon-\mu) \right|_{\mu=0}=\frac{1}{36} \left( 2-6 \ln T \right)$, where we have limited our integration to a temperature-dependent region around the Fermi energy.  The entropy can be seen to follow a $-\ln T$ law similarly. Such a picture is only applicable where only one spin-species is near a feature in the density of states and below the temperature at which the other species picks up the peak.  It must also be above the temperature of the metamagnetic transition where interactions dominate the motion of the Fermi level.  The simple fit therefore only works in a window of temperatures.

These effects mean that a comparison of field and temperature scales is subtle.  Interactions and number conservation alter the rate at which field tunes into the peak in the density of states relative to temperature.  This is shown in figure \ref{fig:specheat} where the peak in $C/T$ is plotted for various interaction strengths and band fillings.  We see that small changes in interaction strength can produce a large change in critical field but has no effect on the temperature of the zero-field maximum.   This means that care is needed when identifying zero field features as a function of temperature with low temperature features at a certain field.

\section{Model $\gamma_2$ band}
\label{sec:g2}
\subsection{Constructing the model}

\begin{figure}
\centering
\includegraphics[width=3in]{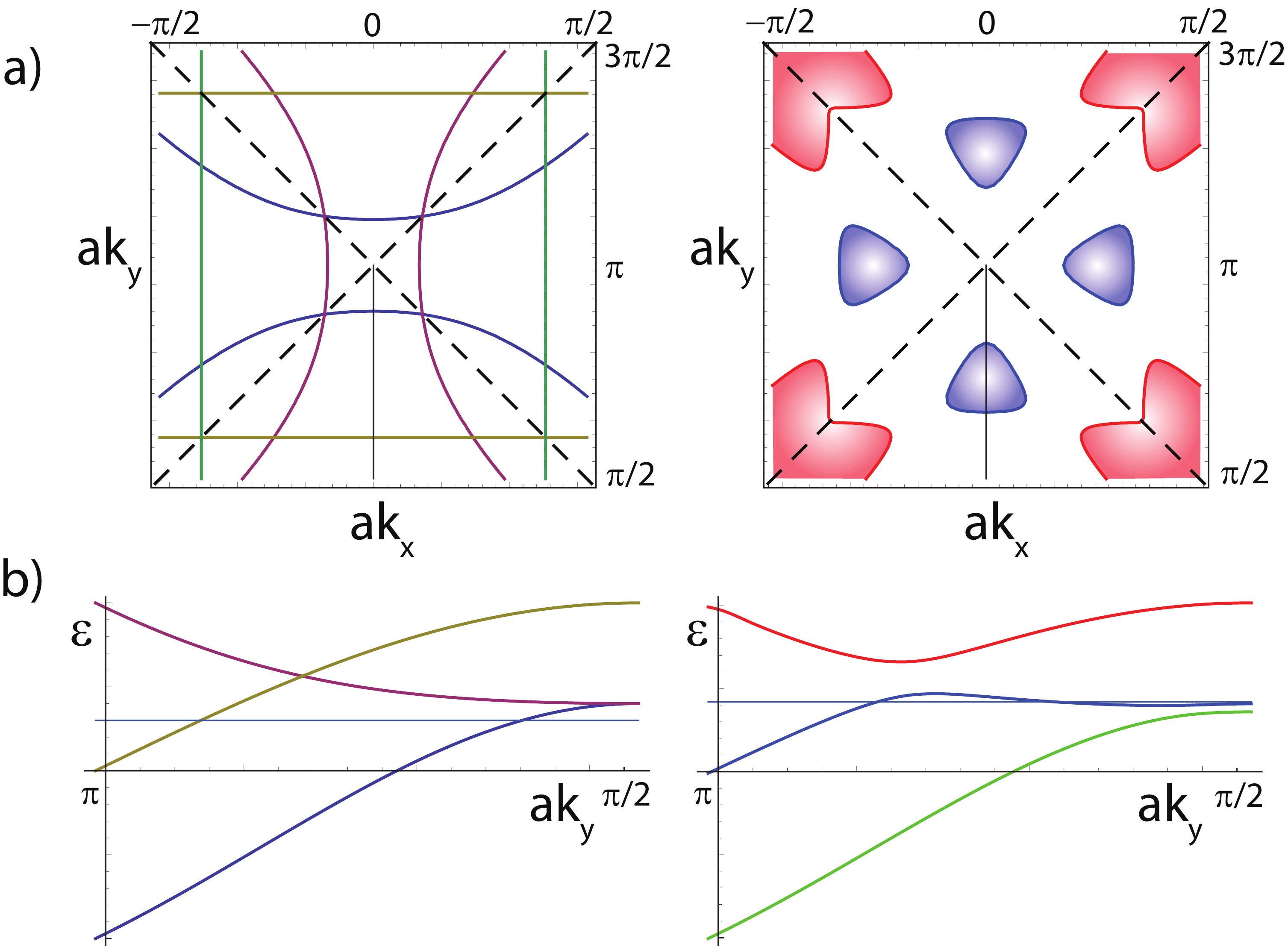}
\caption{\label{fig:gamma2_const} (Color online) a) Fermi surface and dispersions plotted along the indicated line for atomic bands near to the zone corner, $\mu=0.3$.  b) Fermi surface and dispersion for hybridised bands with shift of chemical potential, $\mu=0.42$.  The blue (triangular) hole-like pockets are the bands which we will concentrate on.}
\end{figure}

\begin{figure}
\centering
\includegraphics[width=3in]{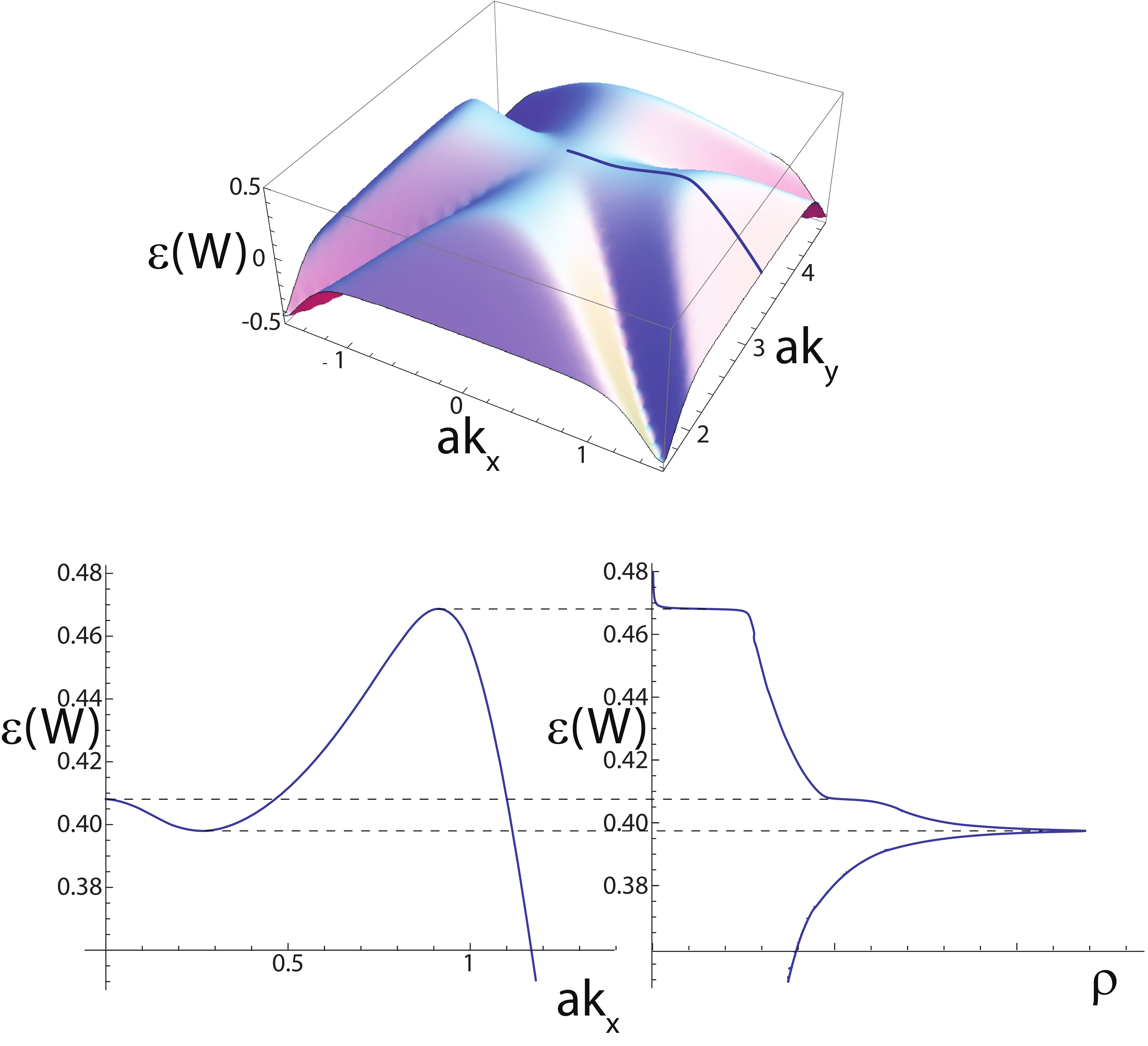}
\caption{\label{fig:gamma2} a) Dispersion for the model $\gamma_2$ pocket.  b) A cut through the dispersion from the zone corner towards the zone center (also indicated in Fig.(\ref{fig:gamma2_const})) showing the saddle point which gives the peak in the density of states and the maximum which gives the shoulder.  The density of states for the model dispersion is shown.  The peak corresponding to the saddle point is clearly visible, as is a step-like feature associated with the local maximum of the dispersion in the zone corner.}
\end{figure}

The metamagnetic transition in \sr\ is thought to be due to a van Hove singularity near to the Fermi surface.  Angle-resolved photoemission spectroscopy (ARPES) studies have identified this as belonging to the electronic band named $\gamma_2$~\cite{327_PRL2008}.  \sr\ is a quasi-two dimensional material with an extrememly complex Fermi surface.  A simplified model of the relevant band can be produced relatively straightforwardly.  We follow a similar but simpler procedure to that in references~\cite{Theory_Kee,Theory_Wu}, identifying a minimal model for the $\gamma_2$ band.  The majority of the Fermi surface in \sr\ is built up from the $t_{2g}$ orbitals of the ruthenium atoms, hybridising via the oxygen $p$-orbitals.  When bilayer splitting and backfolding due to the structural rotation of ruthenium-oxide octrahedra is taken into account this can account for the observed bandstructure~\cite{327_JF_PRB}.  The simplest model for $\gamma_2$ is therefore one involving the $d_{xy}$, backfolded $d_{xy}$ and the $d_{yz}$ and $d_{zx}$ orbitals closest to the Brillouin zone corner as described in appendix \ref{g2model}.  These bands, and the result of including hybridisation between them are shown in figure \ref{fig:gamma2_const} close to the zone corner.  The resulting model for the $\gamma_2$ pocket is shown in figure \ref{fig:gamma2}.  Four triangular pockets form a cross around the zone corner.  A saddle point in the dispersion creates a logarithmic peak in the density of states but additionally there is a shoulder in the density of states due to a local maximum of the band structure in the zone corner.  This feature has consequences for the observable properties of the system.

\subsection{Metamagnetism and Fermi surface transitions}

\begin{figure}
\centering
\includegraphics[width=3in]{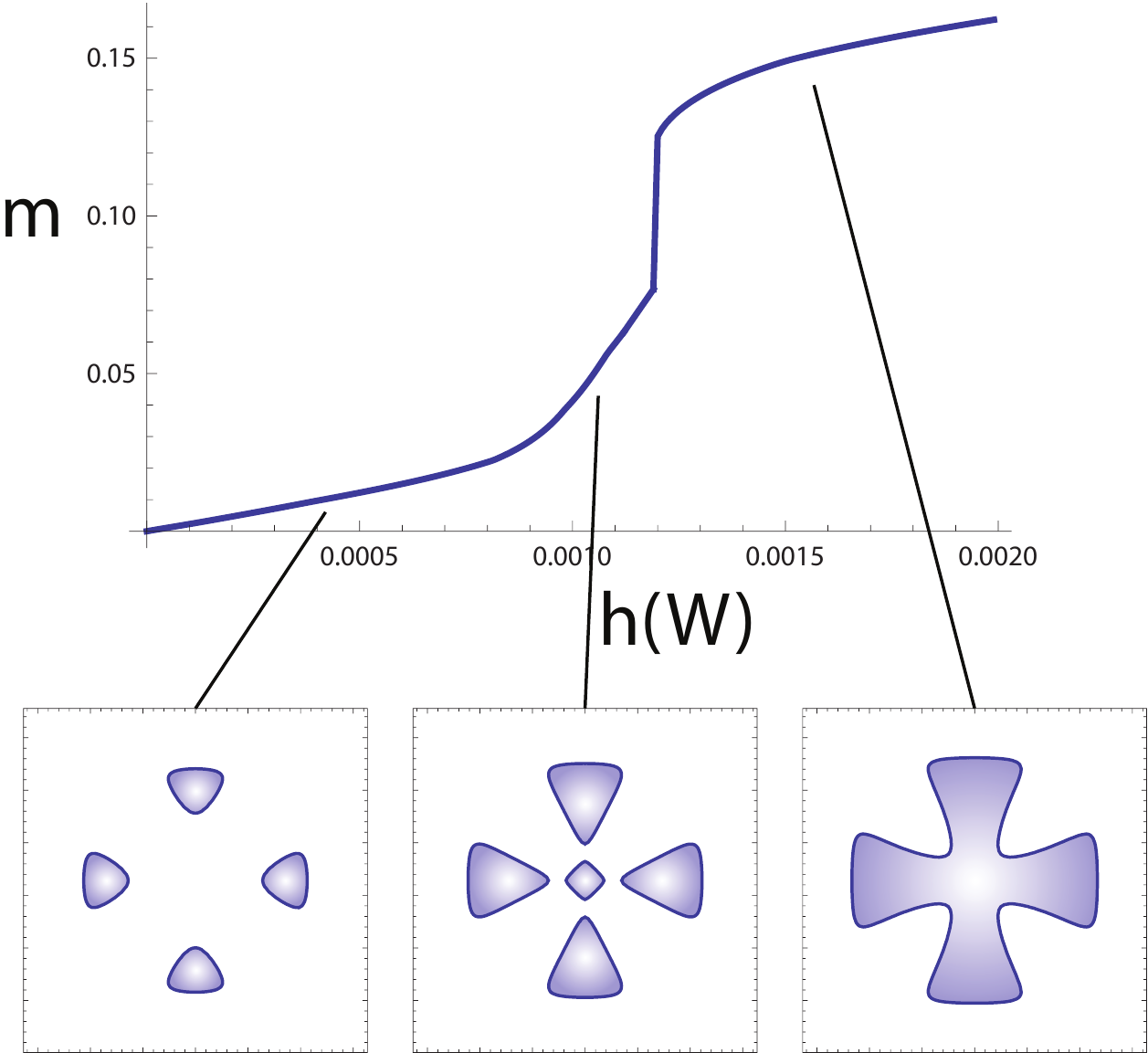}
\caption{\label{fig:gamma2mag}Magnetisation as a function of field for the model $\gamma_2$ pocket, visible are the metamagnetic transition and the magnetic crossover.  The Fermi surfaces for the minority spin-species show the three regimes, low-field, crossover, high-field, as the four pockets connect.}
\end{figure}

The density of states for this model may be straightforwardly used in Eq.(\ref{tm}) to determine the magnetisation profile of the band, with the first order transition determined by minimisation of Eq.(\ref{fenergy}).  This magnetisation is shown in figures \ref{fig:gamma2mag} and \ref{fig:gamma2sp}.  We see that the peak in the density of states reproduces the metamagnetic transition as in the previous case, with a first-order transition which curves towards lower fields as the temperature is increased before becoming continuous at a critical endpoint.  In addition the local maximum of the bandstructure in the zone corner creates a crossover feature - a large but continuous increase in the magnetisation just before the transition.  These features merge into one smooth crossover above the critical point.

The changes in the minority Fermi surface across this transition are plotted in figure \ref{fig:gamma2mag}.  At low fields the four $\gamma_2$ pockets are clearly seperated.  In the crossover regime an additional small pocket appears in the zone center at a Lifshitz transition.  After the metamagnetic transition all of these pockets have reconnected to form a cross-shape.

\subsection{Specific heat and entropy}

\begin{figure*}
\centering
\includegraphics[width=5in]{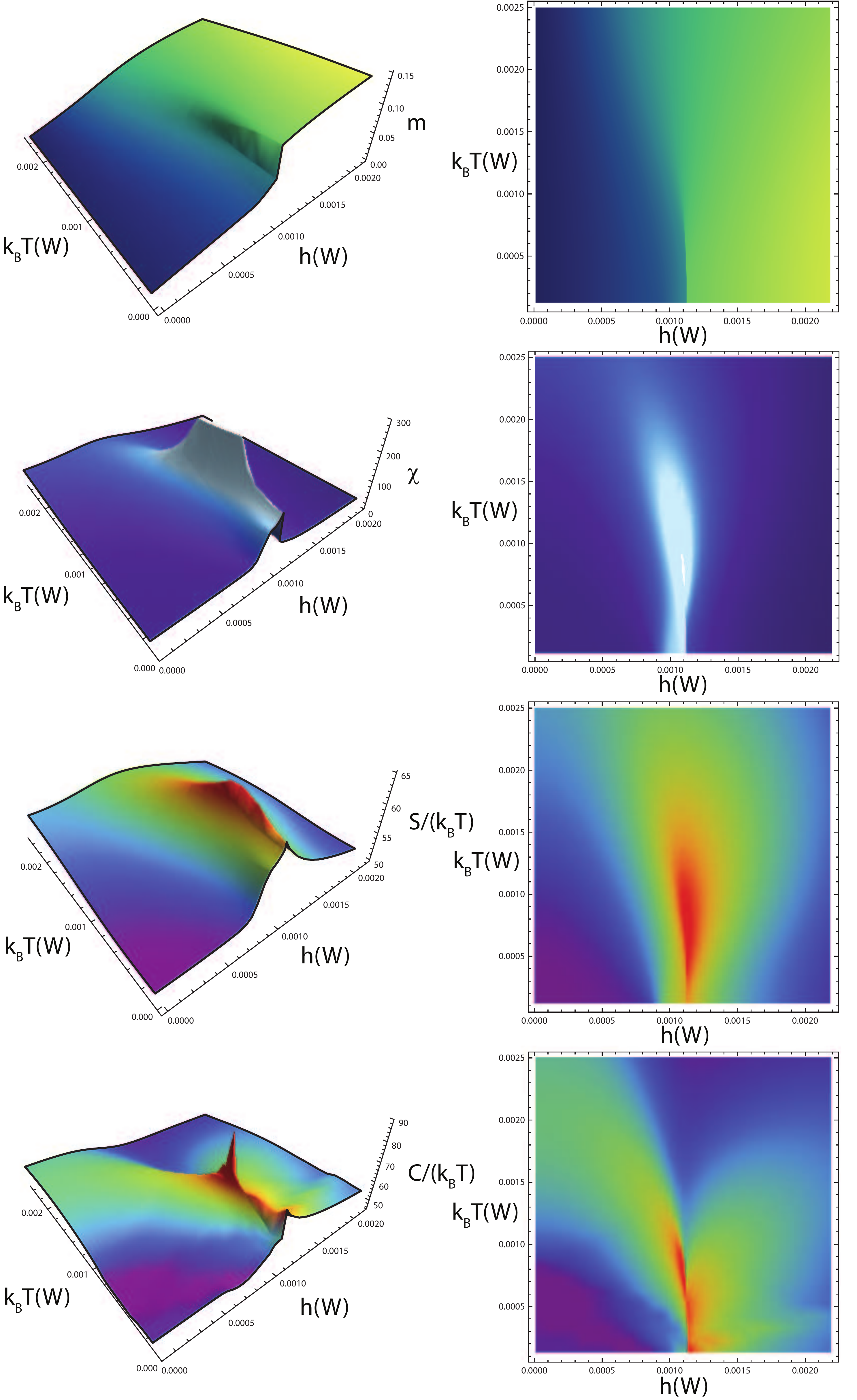}
\caption{\label{fig:gamma2sp} (Color online) a) Magnetisation, b) magnetic susceptibility, c) specific heat and d) entropy as a function of field and temperature for the model $\gamma_2$ pocket.  These are shown as 3D plots and gradient plots to clearly show the temperature dependence.  The first-order transition shows the same behaviour as for the generic logarithmic case but the crossover produces additional shoulder features.  The susceptibility is strongly peaked around the critical endpoint with a weaker feature at the crossover.  Note that the high-temperature shoulder in $C/T$ is present but outwith the temperature range of these plots.  The parameters used for these plots are $g=0.115$, $\mu-\mu_c=0.008$ where $\mu_c$ is the chemical potential of the peak and the density of states is shown in Fig.\ref{fig:gamma2}.}
\end{figure*}

The specific heat and entropy for this pocket may be calculated by substituting the calculated density of states and magnetization into (\ref{sac}) and (\ref{dmu}).  These results are shown in figure \ref{fig:gamma2sp}.  These plots have the same basic form as for the logarithmic density of states (section \ref{sec:results} and Ref.\onlinecite{CS1}) with additional features caused by the crossover.  The first-order transition produces a discontinuity which terminates at the critical endpoint, this discontinuity is in a temperature broadened peak for the entropy or asymmetric double peak for the specific heat.  There is an additional shoulder in $C/T$ moving to higher temperature with field as in the previous case.  The crossover generically produces shoulders in the entropy and specific heat just before the transition as a function of field as one Fermi surface reaches the shoulder in the density of states.  

\begin{figure}
\centering
\includegraphics[width=2.65in]{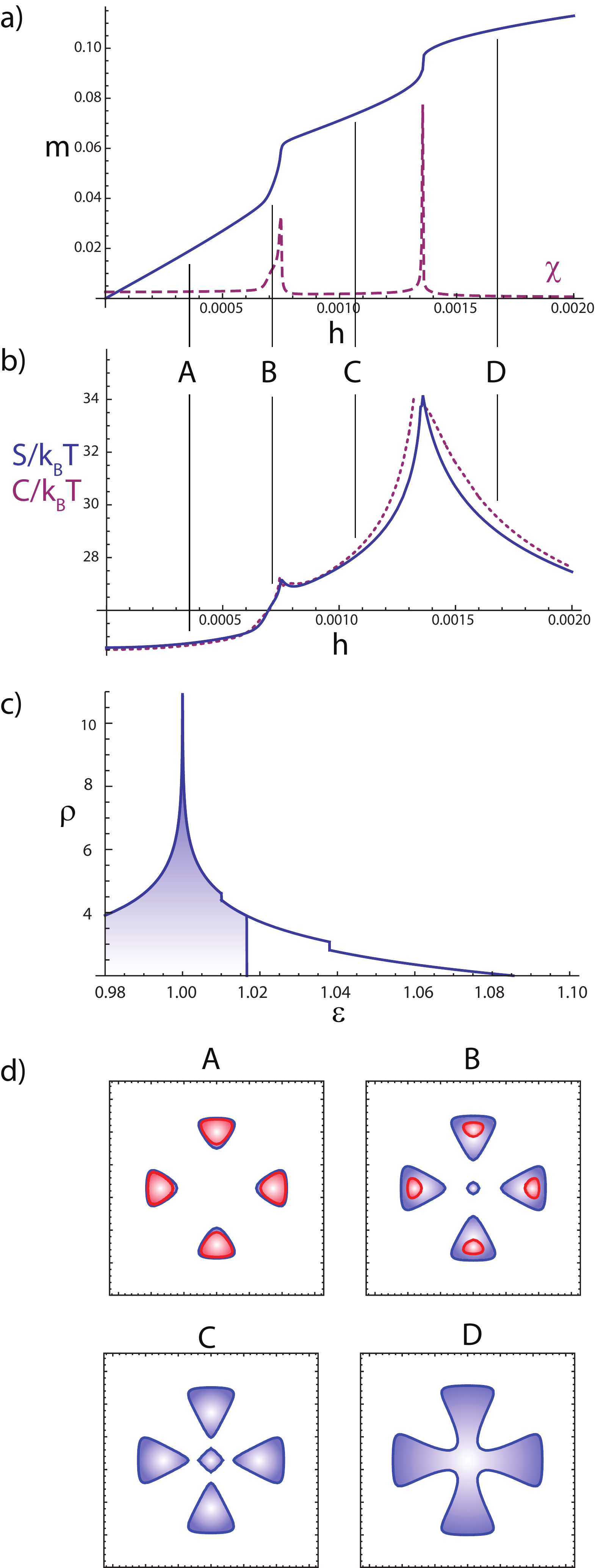}
\caption{\label{fig:bandvanish}(Color online) a) Magnetisation (solid line) and magnetic susceptibility (dashed line) for a model where the crossover is associated with the appearance of a new pocket and the subsequent disappearance of one spin-species.  Letters label distinct Fermi surface topologies.  b) Entropy (solid line) and specific heat (dashed line) for this model, the crossover now produces a peak in the entropy.  c)  The density of states used to produce this behaviour.  The logarithmic singularity produces the first-order transition while the two step features representing the appearance of Fermi surface pockets give the crossover.  Shading represents the zero-field filling.  c) Schematic evolution of the Fermi surface presented earlier through the transitions.  Note the disappearance of one spin species between B and C.  The complete field sweep involves three Lifshitz transitions.}
\end{figure}

While capturing the major features of \sr~\cite{327_Science2009} this model does not match all of of the observed details.  For example the crossover is associated with a distinct additional peak in the entropy in experiment, which is always a shoulder in this picture.  This discrepancy may be taken into account if the majority spin pocket vanishes in the crossover region, giving two Lifshitz transitions in close succession~\cite{*[{}][{ private communication.}] pc_andreas}.  However this requires considerable fine-tuning of the system parameters.  We present a model in which this happens in Fig.\ref{fig:bandvanish}.  In this model we see a kink in the crossover magnetisation associated with a distinct peak in the entropy.  We stress that although the shoulder is a generic feature of this sort of bandstructure this peak requires a density of states and filling and interaction parameters with considerable fine-tuning.

We now consider the field and temperature dependence of the entropy and specific heat.  As discussed in the previous section the van Hove singularity gives a $-\ln{T}$ dependence of $C/T$ above the transition, as observed in \sr~\cite{327_Science2009}.  Number conservation and interactions will alter the field dependence as shown in the previous section.  Additional features of the density of states will also create additional features in the specific heat as a function of temperature and field.  The presence of the shoulder in the density of states produces a steeper rise of the entropy and specific heat as a function of field than would be expected from a van Hove singularity as can be seen in Fig.\ref{fig:gamma2sp}.  This is consistent with experiments where an enhanced divergence was observed~\cite{327_Science2009}.  However in this picture the `divergence' is caused by a feature at a specific energy in the density of states, and its effects will therefore only be seen in a field range with a width proportional to the temperature around this point.  In particular this will result in a very asymmetric dependence of the entropy and specific heat around the van Hove singularity.

We have shown that a simple model for one band of \sr\ can reproduce many of the observed experimental properties. These include non Fermi liquid dependencies as a function of $T$ and $h$, the magnetic crossover and a more rapid divergence of $S$ and $C$ than may be expected from the form of the van Hove singularity.

\section{High entropy regions}
\label{sec:double}

\begin{figure}
\includegraphics[width=3in]{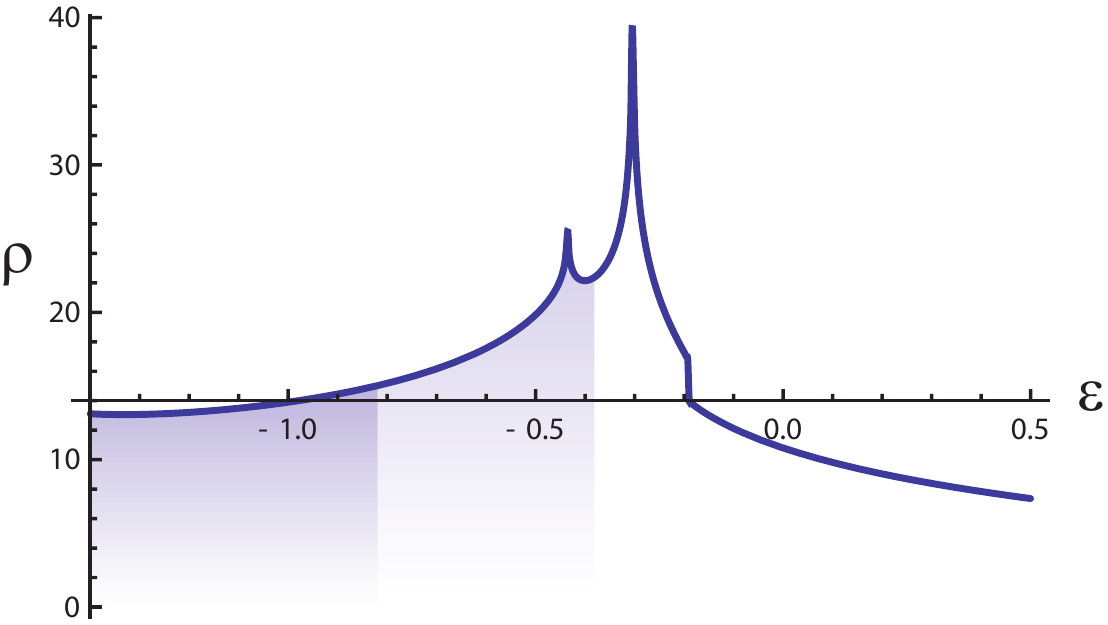}
\caption{\label{fig:sdwdos} Double-peaked density of states obtained from the spin-density wave dispersion.  Shading indicates the Fermi surfaces of the two spin-species in the high-entropy phase.}
\end{figure}

\begin{figure*}
\includegraphics[width=7in]{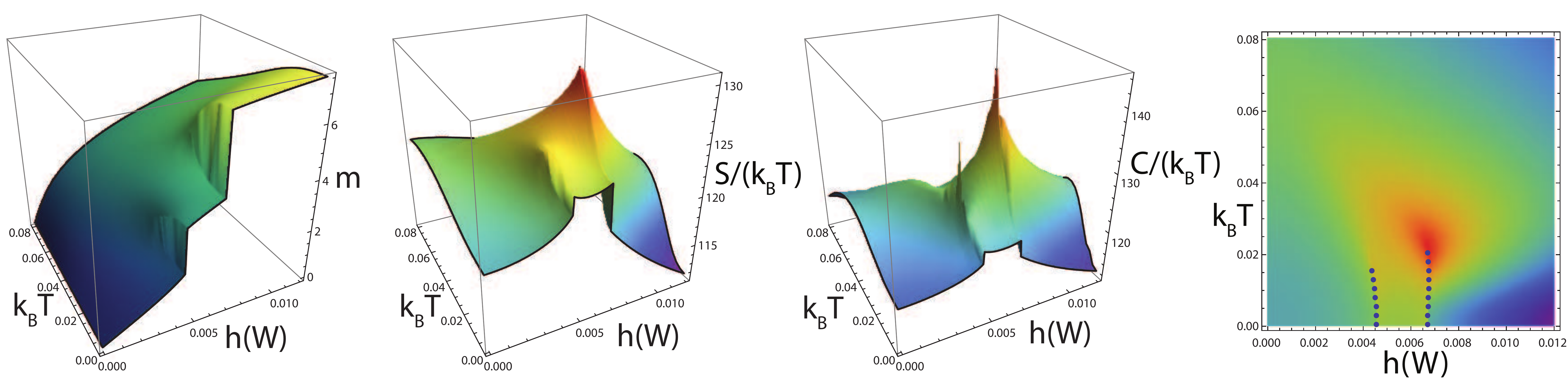}
\caption{\label{fig:sdwent} (Color online) Magnetisation, entropy and specific heat for the density of states shown in Fig.\ref{fig:sdwdos}.  The intermediate magnetisation region and high entropy plateau as a function of field are clearly visible.  The density plot shows S/T with the first-order phase transitions marked.}
\end{figure*}

So far absent from our discussion of \sr\ has been any indication of the anomalous phase.  In the context of our present study the most interesting property of this phase is that upon entering it the entropy jumps to a higher value and upon leaving the entropy falls again.  This produces a region of high entropy with the associated outwards curvature of the first-order phase transitions, which has so-far not been captured in any theoretical treatment.  Examining the entropy jump at the first-order transition in the Stoner theory of the metamagnetic transition suggests that this behaviour may be captured in our picture if one spin-species Fermi surface can be maintained in a region of high density of states after the transition.  This may happen if, for example, we have a double-peaked density of states where the region between the peaks has a higher density of states than that outside.  We now present a theory for the magnetic transitions in such a density of states.  We will find that there are two transitions at low temperature as the Fermi-surface jumps over the two peaks.  In between these transitions there is a plateau which has a higher entropy than the surrounding areas.  In accordance with the Clausius-Clapeyron relation these transitions slope away from each other as a function of temperature.  The density of states in question is shown in Fig.(\ref{fig:sdwdos}) and the results for magnetisation, entropy and specific heat, obtained by the same methods as previously, are shown in Fig.\ref{fig:sdwent}.

Unlike in \sr\ there is no distinct phase between the transitions of this model.  Above the critical endpoint of the transitions is just a crossover, rather than a second-order transition.  Since there is no symmetry broken phase there is no obvious reason for this area to show the resistive anisotropy characteristic of the anomalous phase.  However this approach does show that the thermodynamics of the phase can be captured in a simple mean-field picture.  What is required is a reason for the formation of such a double-peaked density of states.  While we  cannot rule out a simple feature in the density of states, caused by bilayer splitting of the $\gamma_2$ band for example, we point out that this scenario is sensitively dependent on the positions if the peaks.  Small changes in peak height or seperation can destroy the second first-order transition and this configuration is unlikely to have occured by chance.  This may be an interaction-driven effect where the appearance of a new phase reconstructs the density of states.  It is possible that this could be due to the formation of a symmetry broken state such as a spin-density wave, or an electronic nematic state.  In fact the density of states used to calculate the results of figure \ref{fig:sdwent} is that produced from a spin-density wave dispersion with wavevector ${\bf q}$ and magnitude $M$:
\begin{eqnarray}
2E
&=&
\left(\epsilon_{{\bf k}+\frac{\bf{q}}{2}}+\epsilon_{{\bf k}-\frac{\bf{q}}{2}}\right)
+\sqrt{\left(\epsilon_{{\bf k}+\frac{\bf{q}}{2}}-\epsilon_{{\bf k}-\frac{\bf{q}}{2}}\right)^2+(gM)^2},
\nonumber\\
\epsilon_{\bf k}
&=&
-\left(\cos{k_x}+\cos{k_y}\right)+0.3 \cos{k_x} \cos{k_y},
\end{eqnarray}
although our calculation only examines the possibility of ferromagnetic states within this density of states.

\section{Conclusions}

We have shown that a peak in the electronic density of states produces a metamagnetic transition as a function of magnetic field and calculated the corresponding thermodynamic behaviour.  Both the entropy and specific heat show distinctive behaviour.  The entropy has a peak which reflects the density of states but with modified field dependence and temperature broadening.  The transition causes a discontinuous jump in the entropy.  The specific heat also shows a discontinuity but in an underlying double-peak structure.  Such a peak in the density of states arises in the $\gamma_2$ band of \sr\ and is therefore a good candidate for the cause of the metamagnetic transition in this material.  A shoulder in the density of states of this band produces the crossover feature which has been observed in experiment.   The temperature dependence of the specific heat above the transition is logarithmic.  In addition the field dependence is complex, depending on many parameters and not just the density of states.  This raises the question of how to distinguish between deviations from the expected Fermi-liquid behaviour which are due to features in the bandstructure and which are due to the breakdown of the quasiparticle picture - such as at a quantum critical point.

Throughout we have stressed the role of number conservation which can have a significant effect on the critical field of the transition and the field dependence of the other system properties.

Similar models to that presented here have been used to study the bandstructure of \sr~\cite{Theory_orbital,Theory_Kee,Theory_Wu,Theory_Wu_2}.  Many of these include the effects of field angle by introducing spin-orbit and orbital Zeeman coupling, these terms will also cause some change in the bandstructure with field magnitude.  These effects may be straightforwardly introduced into our calculations in the form of a field-dependent density of states, though for simplicity we do not consider this here.

These results provide a strong test for the scenario of a static density of states peak with Zeeman splitting.  In particular the appearance of two features in $C/T$ as a function of temperature at finite magnetic fields is a consequence of this picture.  Careful comparison with experimental measurements on \sr\ may allow the confirmation of the energetic drive behind the metamagnetism and possibly the anomalous phase formation.

In addition to the single metamagnetic transition we have studied the possibility of having phases with a higher entropy than their surroundings in this picture.  We find that a double-peaked density of states can produce this effect and that its properties match well with the thermodynamic signatures of the anomalous phase in \sr, although in this model it is not a distinct phase.  However the underlying reason for such a density of states appearing is outwith the scope of this work.

\acknowledgements{We acknowledge useful discussions with A.G. Green, S.A. Grigera, J.-F. Mercure, A.W. Rost and A.J. Schofield.}

\appendix
\section{Thermodynamic derivations}
\label{thermoderiv}

\subsection{Number conservation}

Enforcing conservation of total number produces a number of relationships between the spin-species which are important for our calculations.  Using the equations and relationships from Section \ref{deriv} we have:
\begin{eqnarray}
\partial_T n 
&=&
\dnua+\dnda
=0
\nonumber\\
\Rightarrow \dnua
&=&
-\dnda,
\nonumber\\
\partial_T m
&=&
\frac{1}{2} \left(\dnua-\dnda\right)
\nonumber\\
&=&\dnua=-\dnda,
\nonumber\\
\partial_T h
&=& \dmua-\dmda-2g\partial_T m
\nonumber\\
&=& \dmua-\dmda-2g\dnua
=
0
\nonumber\\
\Rightarrow \dmua
&=&
\dmda + 2 g \dnua.
\label{numberconservation}
\end{eqnarray}

\subsection{Useful derivatives}

We now evaluate some derivatives which will prove to be useful.  We will use the shortened form $f_\sigma=f(\epsilon-\mu_\sigma)$.
\begin{eqnarray}
\partial_T f_\sigma
&=&
\frac{e^{\frac{(\epsilon-\mags)}{k_B T}}}{(1+e^{\frac{(\epsilon-\mags)}{k_B T}})^2} \left[\frac{\epsilon-\mags}{k_B T^2}+\frac{\dms}{k_B T}\right],
\end{eqnarray}
\begin{eqnarray}
\partial_T \lns
&=&
\frac{1}{1+e^{\frac{\epsilon-\mags}{k_B T}}} \left[\frac{\epsilon-\mags}{k_B T^2}+\frac{\dms}{k_B T}\right]
\nonumber\\
&=&
f_\sigma \left[\frac{\epsilon-\mags}{k_B T^2}+\frac{\dms}{k_B T}\right],
\end{eqnarray}

\subsection{Entropy}

Entropy is the first derivative of the free energy with respect to temperature.

\begin{eqnarray}
S=-\partial_T F
&=&
\sum_\sigma \left[k_B \int d\epsilon \ \dos \lns \right.
\nonumber\\
&&+k_B  T \int d\epsilon \ \dos \partial_T \lns 
\nonumber\\
&&\left.- \dms \nums - \mags \dns\right]
\nonumber\\
&&-
g \left( \dnua\nda+\nua\dnda \right)
+
h \partial_T m
\end{eqnarray}
Using (\ref{numberconservation}) and (\ref{tm})
\begin{widetext}
\begin{eqnarray}
\sum_\sigma \left(\mags\dns + g \dms \dnsd\right)-h \partial_T m
=\left(\nua-\nda+2gm-h\right)\dnua
=0,
\nonumber\\
\end{eqnarray}
\end{widetext}
and so
\begin{eqnarray}
S
&=&
\sum_\sigma \left[k_B \int d\epsilon \ \dos \lns \right.
\nonumber\\
&& \left. + k_B T\int d\epsilon \ \dos \partial_T \lns - \dms\nums \right].
\nonumber\\
\end{eqnarray}

\subsection{Specific Heat}

Specific heat is the first derivative of the internal energy with respect to temperature.  It may alternatively be calculated from the second temperature derivative of the free energy.

\begin{eqnarray}
U
&=&
\sum_\sigma \left[\int d\epsilon \ \epsilon \dos f_\sigma\right]
+g \nua\nda-hm.
\end{eqnarray}
\begin{eqnarray}
C
=
\partial_T U
&=&
\sum_\sigma \left[\int d\epsilon \ \epsilon \dos \partial_T f_\sigma\right]
\nonumber\\&&+
g \left(\dnua\nda+\nua\dnda\right)
-
h \partial_T m
\nonumber\\
&=&
\sum_\sigma \left[\int d\epsilon \ \epsilon \dos \partial_T f_\sigma\right]
-
\left(2gm+h\right)\dnua.
\nonumber\\
\end{eqnarray}

\subsection{Derivatives of the chemical potential}

Before we can evaluate the specific heat or entropy we need to calculate expressions for the temperature derivatives of the chemical potentials.  We begin from the dependence of number on temperature:
\begin{eqnarray}
\dns
&=&
\int d\epsilon \ \dos \partial_T f_\sigma
\nonumber\\
&=&
\int d\epsilon \ \dos \frac{e^{\frac{(\epsilon-\mags)}{k_B T}}}{(1+e^{\frac{(\epsilon-\mags)}{k_B T}})^2} \left[\frac{\epsilon-\mags}{k_B T^2}+\frac{\dms}{k_B T}\right].
\nonumber\\
\label{dtn}
\end{eqnarray}
From (\ref{numberconservation}) and (\ref{dtn}) we have
\begin{eqnarray}
\dms
&=&
-k_B T\frac{\sum_\sigma \int d\epsilon \ \dos \Xi_\sigma \left(\frac{\epsilon-\mags}{k_B T^2}\right)}{\int d\epsilon \ \dos \Xi_\sigma}
\nonumber\\
&&-
\dmsd \frac{\int d\epsilon \ \dos \Xi_{-\sigma}}{\int d\epsilon \ \dos \Xi_\sigma},
\end{eqnarray}
where $\Xi_\sigma=\frac{e^{\frac{(\epsilon-\mu_\sigma)}{k_B T}}}{\left(1+e^{\frac{(\epsilon-\mu_\sigma)}{k_B T}}\right)^2}$.  Also
\begin{eqnarray}
\dms
&=&
\dmsd-2g\dnsd
\nonumber\\
&=&
\dmsd\left(1-\frac{2g}{k_B T} \int d\epsilon \ \dos \Xi_{-\sigma}\right)
\nonumber\\
&&-
2g \int d\epsilon \ \dos \Xi_{-\sigma} \left(\frac{\epsilon-\msd}{k_B T^2}\right),
\end{eqnarray}
resulting in
\begin{eqnarray}
\dms
&=&
\frac{
-k_B T\frac{\sum_\sigma \int d\epsilon \ \dos \Xi_\sigma \left(\frac{\epsilon-\mags}{k_B T^2}\right)}{\int d\epsilon \ \dos \Xi_{-\sigma}}
+
2g \int d\epsilon \ \dos \Xi_\sigma \frac{\epsilon-\mags}{k_B T^2}}
{1-\frac{2g}{k_B T}\int d\epsilon \ \dos \Xi_\sigma+\frac{\int d\epsilon \ \dos \Xi_\sigma}{\int d\epsilon \ \dos \Xi_{-\sigma}}}.
\nonumber\\
\end{eqnarray}

\section{$\gamma_2$ pocket}
\label{g2model}

Our model for the $\gamma_2$ pocket is found by diagonalizing the following Hamiltonian:
\begin{eqnarray}
H &=& \sum_{\sigma,{\bf k}} \psi^\dagger_{\sigma,{\bf k}} \left(\begin{array}{cccc}
d_{xy}\left({\bf k}\right) & \Delta & 0 & \Delta_2\\
\Delta & d'_{xy}\left({\bf k}\right) & \Delta_2 & 0\\
0 & \Delta_2 & d_{zx}\left({\bf k}\right) & 0\\
\Delta_2 & 0 & 0 & d'_{yz}\left({\bf k}\right)\end{array}\right) \psi_{\sigma,{\bf k}},
\nonumber\\
\end{eqnarray}
where $d_{xy}\left({\bf k}\right)$, $d_{yz}\left({\bf k}\right)$, $d_{zx}\left({\bf k}\right)$ are dispersions representing the bands formed by the $t_{2g}$ atomic orbitals:
\begin{eqnarray}
d_{xy}\left({\bf k}\right) &=& -W\left[\cos{\left(k_x\right)}+\cos{\left(k_y\right)}\right]-t \left[\cos{\left(2k_x\right)}+\cos{\left(2k_y\right)}\right]
\nonumber\\
d_{zx}\left({\bf k}\right) &=& -W\cos{\left(k_x\right)}
\nonumber\\
d_{yz}\left({\bf k}\right) &=& -W\cos{\left(k_y\right)},
\end{eqnarray}
and $d'_{xy}$, $d'_{yz}$, $d'_{zx}$ are the backfolded versions with, for example, $d'_{xy}({\bf{k}})=d_{xy}(\bf{k}+\bf{Q})$ where $\bf{Q}=(\pi,\pi)$.
$\Delta$ and $\Delta_2$ are the interaction potentials between bands.  These potentials may arise from spin-orbit effects and therefore allow a coupling of the bandstructure to the angle and strength of the applied field, though for simplicity we do not consider this here.  These bands are shown in figure \ref{fig:gamma2_const} before and after hybridisation with the parameters $t=-0.2W$, $\Delta=0.025W$, $\Delta_2=0.1W$.

\end{document}